\documentclass[sn-mathphys-num]{sn-jnl}

\usepackage{graphicx}%
\usepackage{multirow}%
\usepackage{amsmath,amssymb,amsfonts}%
\usepackage{amsthm}%
\usepackage{mathrsfs}%
\usepackage[title]{appendix}%
\usepackage{xcolor}%
\usepackage{textcomp}%
\usepackage{manyfoot}%
\usepackage{booktabs}%
\usepackage{algorithm}%
\usepackage{algorithmicx}%
\usepackage{algpseudocode}%
\usepackage{listings}%
\usepackage{array}

\usepackage{svg}

\theoremstyle{thmstyleone}%
\theoremstyle{thmstyletwo}%

\theoremstyle{thmstylethree}%

\begin{document}

\title[Decoding TRON]{
Decoding TRON: A Comprehensive Framework for Large-Scale Blockchain Data Extraction and Exploration
}


\author[1]{\fnm{Qian'ang} \sur{Mao}}
\email{me@c0mm4nd.com}

\author[1]{\fnm{Jiaxin} \sur{Wang}}

\author[1]{\fnm{Zhiqi} \sur{Feng}}

\author[1]{\fnm{Yi} \sur{Zhang}}

\author*[1]{\fnm{Jiaqi} \sur{Yan}}

\affil*[1]{\orgdiv{School of Information Management}, \orgname{Nanjing University}, \orgaddress{\street{Xianlin Road}, \city{Nanjing}, \postcode{210023}, \state{Jiangsu}, \country{China}}}




\abstract{
%

Cryptocurrencies and Web3 applications based on blockchain technology have flourished in the blockchain research field. Unlike Bitcoin and Ethereum, due to its unique architectural designs in consensus mechanisms, resource management, and throughput, TRON has developed a more distinctive ecosystem and application scenarios centered around stablecoins. Although it is popular in areas like stablecoin payments and settlement, research on analyzing on-chain data from the TRON blockchain is remarkably scarce. To fill this gap, this paper proposes a comprehensive data extraction and exploration framework for the TRON blockchain. An innovative high-performance ETL system aims to efficiently extract raw on-chain data from TRON, including blocks, transactions, smart contracts, and receipts, establishing a research dataset. An in-depth analysis of the extracted dataset reveals insights into TRON's block generation, transaction trends, the dominance of exchanges, the resource delegation market, smart contract usage patterns, and the central role of the USDT stablecoin. The prominence of gambling applications and potential illicit activities related to USDT is emphasized. The paper discusses opportunities for future research leveraging this dataset, including analysis of delegate services, gambling scenarios, stablecoin activities, and illicit transaction detection. These contributions enhance blockchain data management capabilities and understanding of the rapidly evolving TRON ecosystem.

}




\maketitle

\section{Introduction}

In recent years, the blockchain ecosystem has experienced tremendous growth, as evidenced by the rising market capitalization and the adoption of blockchain-based cryptocurrencies and platforms. Bitcoin, the first and most well-known cryptocurrency, had a market capitalization exceeding \$1.35 trillion in March 2024, surpassing the market value of silver at one point. Bitcoin's success has inspired countless other blockchain projects in areas such as decentralized finance (DeFi), non-fungible tokens (NFTs), and Web3. Similarly, in March 2024, the total valuation of the cryptocurrency market exceeded \$2.24 trillion, indicating the rapid expansion of the blockchain ecosystem.
Reflecting this growth, academic research on blockchain technology has also surged. Early research primarily focused on fundamental blockchain technologies, including consensus algorithms, cryptography, and scalability solutions. As blockchain platforms and applications have become firmly established, recent research attention has gradually shifted towards analyzing and improving the design, applications, and user experience of the blockchain ecosystem. 
For example, active research areas include DeFi protocol analysis, NFT markets, DAO formation and governance, blockchain gaming, and factors influencing user adoption. 
New specialized journals and conferences focusing on blockchain applications and business use cases have also emerged.
While popular blockchain platforms like Bitcoin and Ethereum and their smart contract applications have garnered widespread research attention, other ecosystems with unique architectural designs and use cases remain relatively unexplored. The vast amounts of data generated by these specialized blockchain systems, although possessing significant commercial and academic value similar to Bitcoin and Ethereum, also present substantial technical challenges for analyzing heterogeneous blockchain data.

In 2021, TRON surpassed Ethereum in USDT stablecoin supply, becoming a leading stablecoin issuance platform globally. By 2023, TRON reached 200 million users, with 34.5 million (approximately 17.2\%) holding USDT. However, TRON's heavy focus on stablecoin transactions poses risks, as it was flagged by Israel and the United States for aiding terrorist fundraising activities. TRON's popularity among terrorist groups is attributed to its faster transaction speeds and lower fees compared to Bitcoin, making it a preferred platform for illicit activities like money laundering

However, current research on TRON on-chain data is scarce, with most studies focusing only on the basic mechanism design analysis of the price fluctuations of its native token TRX and USDT stablecoin \cite{yadav2021qualitative,li2020resource,li2023hard,li2024stablecoin,shukla2023trx,maghsoodi2023cryptocurrency}. 
The fundamental challenges stem from several factors. Firstly, there is an absence of a universal data extraction tool for TRON. While certain blockchain websites like TRONSCAN \footnote{\url{https://tronscan.org}} offer partial TRON data, their data extraction tools are not publicly accessible, and the data acquisition process is rate-limited, restricting access to comprehensive datasets. Secondly, there is a lack of comprehensive data exploration tools specifically designed for TRON. Although extensive research has been conducted on data analysis for EOS.IO and Ethereum \cite{zheng2021xblockEOS,zheng2020xblockETH}, studies focusing on TRON are scarce. To the best of our knowledge, there has been no comprehensive analysis performed on the entirety of TRON's dataset across all data types, leaving a significant gap in understanding the intricacies of this blockchain platform. Thirdly, the extraction and processing of TRON's data present significant difficulties. TRON's data types are based on Protocol Buffers, and it employs gRPC to deliver high-performance interfaces, encompassing numerous intricate data structures and data types. Furthermore, the data structures within TRON involve nesting, arbitrary types, and other complex relationships, posing significant challenges for data parsing endeavors and hindering the development of effective analysis tools and techniques.

This paper addresses the challenges and motivations surrounding TRON blockchain data analysis through several key contributions. We present a comprehensive data processing framework for the TRON blockchain, featuring robust ETL workflows and querying capabilities, which enables efficient extraction and analysis of the vast TRON ecosystem data. Our work includes detailed explanations of multiple datasets derived from this processing methodology, accompanied by preliminary analyses to illuminate the data's content and potential applications. Additionally, we provide a critical assessment of the current research landscape and explore promising future directions that could leverage these datasets. By offering these tools and insights, we aim to empower researchers and developers in the blockchain analytics field, fostering innovation and a deeper understanding of the TRON blockchain. This research not only advances the technical capabilities for blockchain data processing but also paves the way for novel applications and scholarly investigations in this rapidly evolving domain.

\section{Background}

\subsection{TRON Consensus}

TRON's consensus mechanism is different from that of Bitcoin and Ethereum, and it adopts the same DPoS (Delegated Proof of Stake) as EOS.io. The DPoS consensus mechanism is a way to verify transactions and maintain network security by electing representative nodes (i.e., SR, Super Representatives). Users vote for 27 Super Representatives by staking (freezing) their held TRX (TRON's cryptocurrency). Super Representatives are responsible for generating blocks and processing transactions on the network and are re-elected every 6 hours to ensure the network's decentralization and efficiency. The DPoS mechanism increases the network's processing speed and transaction throughput by reducing the number of nodes participating in the consensus while incentivizing users to actively participate in network governance.

\subsection{TRON Account Model}

TRON uses an account model. The address is the unique identifier of the account, and operating the account requires a private key signature. The account has many attributes, including TRX and TRC10 token balances, bandwidth, energy, and more. The account can send transactions to increase or decrease its TRX or TRC10 token balance, deploy smart contracts, and trigger its own or others' published smart contracts, which are self-executing codes that automatically execute predetermined actions when specific conditions are met. All TRON accounts can apply to become Super Representatives or vote for elected Super Representatives. The account is the foundation of all activities on TRON.

\subsection{TRON Transaction Execution}

The TRON Transaction Execution is a comprehensive process that begins with transaction creation, typically initiated by a user through a wallet or decentralized application (dApp). Once created, the transaction is broadcast to the TRON network, where it's verified by nodes and enters the mempool (transaction pool) to await processing. SR nodes, key players in TRON's consensus mechanism, select transactions from the pool to package into blocks. For transactions involving smart contracts, the TRON Virtual Machine (TVM) comes into play. The TVM, a core component of the TRON network, is responsible for executing smart contract code. It's Turing-complete and compatible with the Ethereum Virtual Machine (EVM), allowing it to run contracts written in Solidity. The TVM processes the contract by loading its code, interpreting opcodes, executing logic, updating states, and consuming energy (similar to Ethereum's gas). After contract execution, the SR nodes reach consensus on the block's validity using the DPoS mechanism. Once consensus is achieved, the new block is added to the blockchain, confirming the transactions it contains. This process updates the global state of the TRON network, including account balances and contract states. Any events triggered by the transactions or contract executions are recorded, allowing dApps to listen and respond accordingly. The transaction is then considered complete, with its results visible across the network and accessible through block explorers or wallets. Throughout this workflow, the TVM plays a crucial role, particularly in handling smart contract transactions, enabling TRON to support complex decentralized applications and DeFi projects.

\subsection{TRON Resource Model}

TRON's resource model manages transactions and smart contract execution on the network through two resources: ``Bandwidth" and ``Energy".

Bandwidth is the resource consumed by users when performing ordinary transactions (such as transfers), and each account receives a certain amount of free bandwidth every day. Users can obtain additional bandwidth by freezing (locking for a period of time) their held TRX (TRON's cryptocurrency). Freezing TRX not only increases bandwidth but also grants voting rights for network governance.

Energy is the resource consumed by users when executing smart contracts. Like bandwidth, users can obtain energy by freezing TRX. The mechanism of freezing TRX to obtain energy and bandwidth encourages users to actively participate in network activities, providing a decentralized way to allocate and manage computing resources on the blockchain, ensuring the network's security and efficient operation.

\section{Our Framework}




This chapter describes the process of obtaining raw data from the TRON blockchain. The extraction and analysis processes in our framework are described below:

\subsection{Extract, Transform, and Load}

\begin{figure}[htbp]
\includegraphics[width=\textwidth]{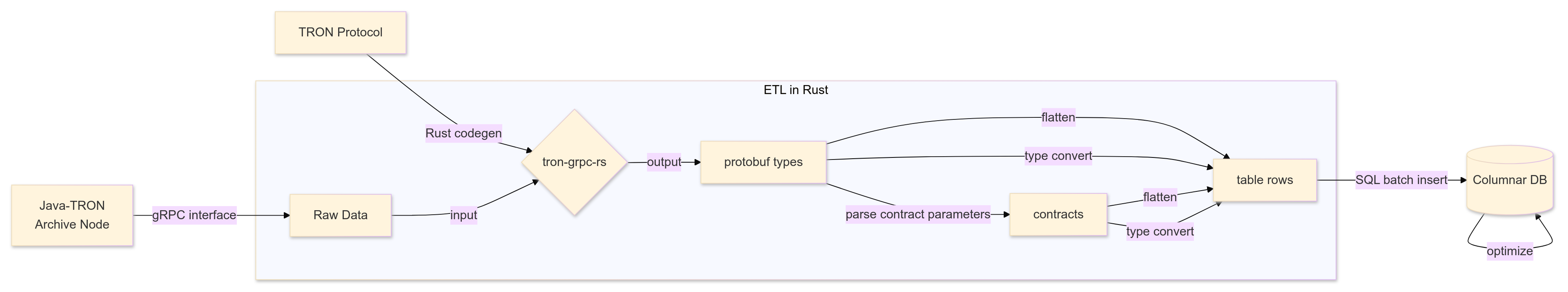}
\caption{This figure illustrates the ETL(Extract, Transform, Load) pipeline for TRON blockchain data, which shows the flow from a Java-TRON Archive Node, through a Rust-based processing system, to final storage in a Columnar Database. 
} \label{fig:extraction}
\end{figure}

\autoref{fig:extraction} illustrates an ETL (Extract, Transform, Load) process implemented in Rust, specifically designed to handle data from the TRON blockchain’s Java-TRON Archive Node. The source code of the ETL is available on our GitHub\footnote{\url{https://github.com/njublockchain/web3research-etl}}. The process begins with the extraction phase, where raw data is retrieved from the Java-TRON Archive Node via a gRPC interface. This raw data is then processed by the tron-grpc-rs library, which is built using Rust and generated from the TRON Protocol’s specifications. This library is responsible for handling the gRPC communication and converting the extracted data into Protobuf types.

Within the transformation phase, encapsulated in a Rust-based ETL submodule, the Protobuf types undergo several operations to prepare the data for storage. First, the data is flattened, transforming complex, nested structures into simple, table-like rows. This is followed by a type conversion step, where data types are adapted to formats suitable for downstream processes. Additionally, the process involves parsing contract parameters from the Protobuf data, which are critical for understanding and processing smart contracts within the TRON ecosystem. The parsed contract parameters are also subjected to flattening and type conversion, ensuring that all data is uniformly structured and ready for database insertion.

The final phase of the ETL process is the loading step, where the processed table rows are batch-inserted into a columnar database via SQL. Post-insertion, the database undergoes an optimization process to enhance data retrieval efficiency and overall performance. This step is crucial for maintaining the scalability and responsiveness of the database when handling large volumes of blockchain data.

\subsection{Data Exploration}

\begin{figure}[htbp]
\includegraphics[width=\textwidth]{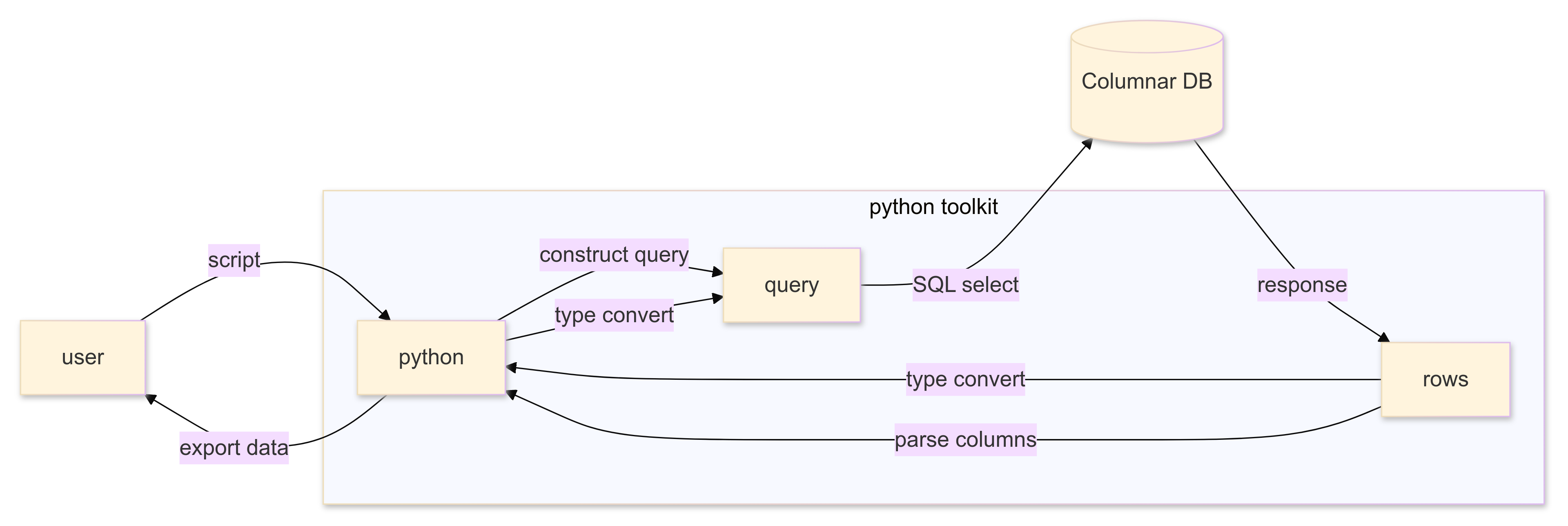}
\caption{This figure illustrates the data exploration process using Python to query a columnar database, which shows how a user's script interacts with Python to construct queries, convert data types, and parse results from a columnar database, with the option to export data back to the user.} \label{fig:exploration}
\end{figure}

\autoref{fig:exploration} presents a detailed exploration process, illustrating the interaction between the user, a Python toolkit, and a columnar database in the context of data analysis. The source codes of the Python toolkit and the corresponding columnar database DDL(Data Definition Language) are available on our GitHub\footnote{\url{https://github.com/njublockchain/web3research-py}}. Initially, the user engages with the Python environment through a script, which acts as a conduit for sending instructions to the Python toolkit. Upon receiving these instructions, the Python toolkit initiates a sequence of processes designed to facilitate data retrieval and manipulation.

The first major step involves the construction of a query, wherein the Python toolkit interprets the user’s instructions and formulates an appropriate database query. This query formulation process may require type conversion to ensure that the data types align with the expected formats in the database. The query, once constructed, is then executed against a columnar database via an SQL SELECT statement. This database, specialized in handling columnar data storage, processes the query and returns the requested data organized into rows.

Following the retrieval of data, the Python toolkit engages in further processing steps. These include another round of type conversion to adapt the retrieved data to a format suitable for subsequent analysis and a parsing operation to extract and organize the relevant columns from the dataset. Once these processes are complete, the toolkit exports the processed data back to the user, thereby completing the data exploration cycle.

\section{Datasets}
In this section, we will provide a detailed description of the statistics and observations of several datasets obtained from the TRON blockchain through the processing of the above framework.

\begin{table}
\centering
\small
\begin{tabular}{@{}ccr@{}}
\toprule
\textbf{Main Category} & \textbf{Subcategory} & \textbf{Contracts} \\
\midrule
\multirow{13}{*}{Assets} & \multirow{3}{*}{TRC10} & assetIssueContracts \\
 &  & participateAssetIssueContracts \\
 &  & updateAssetContracts \\
\cmidrule(l){2-3}
& \multirow{3}{*}{Transfer} & transferContracts \\
 & & transferAssetContracts \\
 & & shieldedTransferContracts \\
\cmidrule(l){2-3}
& \multirow{7}{*}{Staking} & cancelAllUnfreezeV2Contracts \\
 & & freezeBalanceContracts \\
 & & freezeBalanceV2Contracts \\
 & & unfreezeAssetContracts \\
 & & unfreezeBalanceContracts \\
 & & unfreezeBalanceV2Contracts \\
 & & withdrawBalanceContracts \\
\midrule
\multirow{14}{*}{Account} & \multirow{5}{*}{EOAs} & accountCreateContracts \\
 & & accountPermissionUpdateContracts \\
 & & accountUpdateContracts \\
 & & setAccountIdContracts \\
 & & updateSettingContracts \\
\cmidrule(l){2-3}
 & \multirow{3}{*}{SmartContracts} & clearAbiContracts \\
 & & createSmartContracts \\
 & & triggerSmartContracts \\
\cmidrule(l){2-3}
& \multirow{5}{*}{Resource} & delegateResourceContracts \\
 & & undelegateResourceContracts \\
 & & updateBrokerageContracts \\
 & & updateEnergyLimitContracts \\
 & & withdrawExpireUnfreezeContracts \\
\midrule
\multirow{8}{*}{DEX} & \multirow{4}{*}{Exchange} & exchangeCreateContracts \\
 & & exchangeInjectContracts \\
 & & exchangeTransactionContracts \\
 & & exchangeWithdrawContracts \\
\cmidrule(l){2-3}
 & \multirow{2}{*}{Market} & marketCancelOrderContracts \\
 & & marketSellAssetContracts \\
\midrule
\multirow{8}{*}{Government} & \multirow{3}{*}{Proposal} & proposalApproveContracts \\
 & & proposalCreateContracts \\
 & & proposalDeleteContracts \\
\cmidrule(l){2-3}
 & \multirow{4}{*}{SR Voting} & voteWitnessContracts \\
 & & voteAssetContracts \\
 & & witnessCreateContracts \\
 & & witnessUpdateContracts \\
\bottomrule
\end{tabular}
\caption{The classification table for our parsed dataset of TRON protocol contracts. We grouped the data by primary use case into four categories: Assets, Accounts, DEXs (Decentralized Exchanges), and Governance. Each category is further divided by specific function: Assets into TRC10, Transfer, and Staking; Accounts into EOAs (Externally Owned Accounts), Smart Contracts, and Resources; DEXs into Exchanges and Markets; and Governance into Proposals and SR (Super Representative) Voting.}
\label{tab:protocol_contracts}
\end{table}

\autoref{tab:protocol_contracts} shows the classification relationship from the raw data to the seven data sets.
We collect data block chain chain will be accessible from the public platform\footnote{\url{https://web3resear.ch/datasets}}. 

\subsection{Blocks}

Blocks, as the name implies, are essential components of the blockchain data structure, serving as packages of transactions. In our Blocks dataset, we primarily retain Block Header information, while specific transaction details are stored in an external transaction dataset.

The core fields in this dataset include block hash, timestamp, Merkle root hash of transactions, parent block hash, block height, witness ID and address, block version number, account state tree root hash, witness signature, and transaction count. The block hash serves as a unique identifier for each block, ensuring the integrity and immutability of block data. The timestamp field records the precise time of block generation, facilitating research on block production rates and temporal distribution. The Merkle root hash of transactions and the account state tree root hash are used to verify transactions and account states within the block, respectively, reflecting TRON network's design in data validation and state management. The parent block hash maintains the linkage between blocks, guaranteeing the continuity and consistency of the blockchain. The block height field indicates the position of the block within the entire chain, serving as a crucial parameter for analyzing network growth and historical queries.

Witness-related fields, such as witness ID, address, and signature, directly reflect the characteristics of TRON's DPoS consensus mechanism. This information not only validates the legitimacy of blocks but also enables analysis of Super Representative activity patterns and network participation. The block version number field reflects the evolution of the TRON protocol, aiding in tracking network upgrades and compatibility changes. The transaction count field records the number of transactions in each block, providing significant data for analyzing network throughput, user activity, and economic activity scale.

Through in-depth analysis of this comprehensive block dataset, researchers can gain insights into TRON network's operational characteristics, performance metrics, security, and degree of decentralization. For instance, timestamp and block height data can be used to study network stability and throughput variations; witness-related data can be analyzed to examine Super Representative election mechanisms and power distribution; transaction count and block size can be utilized to explore network scalability and user adoption trends.

\subsection{External Transactions}

In TRON, external transactions, similar to those in Ethereum, represent transactions initiated by EOA (Externally Owned Account) addresses. This dataset focuses on external transaction data within the TRON blockchain network, encompassing not only initial transaction information but also merged data on post-execution results. This design provides researchers with a complete view of a transaction's entire lifecycle, from initiation to final on-chain confirmation. Each record represents an independent external transaction, containing both the original transaction data and detailed results of its execution on the TRON network.

The core fields of the dataset can be broadly categorized into several key areas: transaction identification information, block information, authorization data, contract call data, resource consumption and fee information, execution results, and special operation data. The transaction hash serves as a unique identifier for each transaction, ensuring traceability. Block number and transaction index precisely locate the transaction within the blockchain, crucial for studying transaction temporality and block filling efficiency.

Authorization-related fields (such as authorityAccountNames, authorityAccountAddresses, etc.) reflect TRON network's multi-signature and complex permission management mechanisms. These data provide a foundation for analyzing network security models and user interaction patterns. Contract-related fields (contractType, contractParameter, etc.) detail smart contract invocations, significant for understanding the usage patterns of decentralized applications (DApps) and the popularity of smart contracts in the TRON ecosystem. Furthermore, we have parsed contractParameters into multiple sub-datasets based on different contractTypes, which will be introduced in the next section.

Notably, this dataset incorporates transaction execution result data. Fields such as energyUsage, netUsage, and fee meticulously record resource consumption, crucial for analyzing TRON network's resource pricing mechanisms and efficiency. The receiptResult and result fields directly reflect transaction execution outcomes, while the resMessage field may contain more detailed execution information or error descriptions. The combination of these data enables researchers to conduct in-depth analyses of transaction failure reasons, network congestion situations, and smart contract execution efficiency.

The dataset also includes dedicated fields for special operations, such as asset issuance (assetIssueId), account freezing and unfreezing (withdrawAmount, unfreezeAmount), exchange operations (exchangeId, exchangeReceivedAmount, etc.), and order-related information (orderId, orderDetails, etc.). These fields reflect the diverse financial functionalities supported by the TRON network, providing direct data support for researching decentralized finance (DeFi) activities on the network.

Through comprehensive analysis of this external transaction dataset, researchers can gain multi-faceted insights into TRON network operations. For instance, they can study the usage frequency, resource consumption patterns, and success rates of different types of smart contracts, evaluating network resource utilization efficiency and the rationality of pricing mechanisms. Transaction result and resource usage data can be used to analyze network performance bottlenecks and optimization opportunities. Authorization account and signature information can be employed to study network security and user behavior patterns. Data on special transaction types provides valuable empirical foundations for researching financial innovations and user adoption within the TRON ecosystem.

\subsubsection{Protocol Contracts}

\subsection{Smart Contract Event Logs}

During the transaction execution process in TRON, smart contracts on the TVM (TRON Virtual Machine) also generate a special data structure called Event Log, used to record events and log information during contract runtime. Event Log is essentially a data structure actively triggered and defined by smart contract code, allowing developers to record custom data at critical execution points, such as state changes, error messages, audit trails, etc. The dataset used in this study focuses on event log data from the TRON blockchain network, capturing various events generated during smart contract execution in the TVM. In the blockchain ecosystem, event logs play a crucial role, not only recording key points of contract execution but also providing an important interface for off-chain applications to interact with on-chain activities. The unique value of this dataset lies in its detailed record of smart contract activities on the TRON network, offering valuable insights for researchers to deeply analyze contract behaviors, user interaction patterns, and the dynamics of the entire ecosystem.

The core fields of the dataset include block number, associated transaction hash, log index, contract address, up to four topic fields (topic0 to topic3), and a data field. Each record represents an independent event log, precisely located within a specific block and transaction, and distinguished by the log index for multiple events within the same transaction.

The block number (blockNum) and transaction hash (transactionHash) fields link the event logs to the overall structure of the blockchain, allowing researchers to track the position and timing of events throughout the blockchain's history. The log index (logIndex) further refines the order of multiple events within the same transaction, which is crucial for understanding the execution flow and results of complex transactions.

The contract address (address) field identifies the smart contract that triggered the event, providing a basis for analyzing activity patterns and user interaction frequencies of specific contracts. The topic fields (topic0 to topic3) are the indexed part of the event, typically containing the event signature (topic0) and key parameters. The nullable design of these fields increases the flexibility of the data structure, accommodating the diverse needs of different types of events. The data field contains the non-indexed part of the event, potentially including more detailed parameter information.

Through in-depth analysis of this event log dataset, researchers can gain multifaceted insights into the smart contract ecosystem of the TRON network. For example, they can study the frequency and distribution of specific types of events, identify the most active contracts and the most common user interaction patterns. Combined with transaction data, a complete smart contract activity map can be constructed, providing a deep understanding of complex DApp operation mechanisms and user behavior patterns.

Moreover, this dataset is particularly valuable for studying the operations of decentralized finance (DeFi) applications. For instance, token transfer events can be analyzed to track fund flows, liquidity provision and removal events can be studied to evaluate the health of DeFi protocols, or price oracle events can be examined to research the propagation and impact of market data on-chain.

Event log data also provides important tools for developers and auditors. By analyzing error events and abnormal patterns, potential contract vulnerabilities or security risks can be identified. Additionally, these data form the basis for building efficient indexing and real-time monitoring systems, supporting more complex off-chain analysis and application development.

\subsection{Internal Transaction}

Similar to Ethereum, in TRON's transaction system, apart from common transactions (also known as External Transactions), there exists a special type of transaction called Internal Transactions. This dataset focuses on internal transaction data within the TRON blockchain network, capturing internal calls and value transfers generated during smart contract execution. Internal transactions differ from external transactions in that they are not directly initiated by users, but are triggered by contract code during smart contract execution. The significance of this dataset lies in its revelation of complex interaction patterns and detailed value flows of smart contracts on the TRON network, providing researchers with a valuable resource for in-depth understanding of the internal operational mechanisms of the TRON ecosystem.

The core fields of the dataset include block number, associated external transaction hash, internal transaction index, internal transaction hash, caller address, recipient address, token transfer information, notes, whether the transaction was rejected, and additional information. Each record represents an independent internal transaction, precisely located within a specific block and external transaction, and distinguished by an internal index for multiple internal calls within the same external transaction.

The block number (blockNum) and external transaction hash (transactionHash) fields link internal transactions to the overall blockchain structure, allowing researchers to track the position and sequence of internal transactions throughout the transaction execution process. The internal transaction index (internalIndex) further refines the order of multiple internal calls within the same external transaction, which is crucial for understanding the execution flow of complex smart contracts.

The caller address (callerAddress) and recipient address (transferToAddress) fields reveal patterns of inter-contract calls and paths of value flow. This data is significant for analyzing smart contract interaction networks, identifying key contracts, and tracing fund flows. Token transfer information (callValueInfos.tokenId and callValueInfos.callValue) is stored in array form, supporting simultaneous transfers of multiple tokens, reflecting the complex financial operations supported by the TRON network.

The field indicating whether a transaction was rejected (rejected) provides direct information about the success or failure of internal transaction execution, which is valuable for assessing the reliability of smart contracts and identifying potential vulnerabilities or design flaws. The note and extra information fields may contain additional explanations about the purpose of internal transactions or special conditions, providing extra context for in-depth analysis.

Through comprehensive analysis of this internal transaction dataset, researchers can gain multi-faceted insights into the smart contract ecosystem of the TRON network. For instance, they can study common contract interaction patterns, identify frequently called contracts, and key value transfer paths. Combined with external transaction data, a complete transaction execution graph can be constructed, enabling deep understanding of complex DApp operational mechanisms. Analysis of the rejected field can help evaluate the robustness of smart contracts and potential security risks. Analysis of token transfer information can reveal token economic activities and liquidity patterns on the TRON network.

Furthermore, this dataset is particularly valuable for studying the internal working mechanisms of decentralized finance (DeFi) applications. For example, it allows analysis of internal transaction processes of complex lending protocols, decentralized exchanges, or cross-chain bridges, providing understanding of how funds flow between different contracts and the resource consumption of various operations.
\section{On-chain Data Insight}

Based on the above data-extracting methodology, we have acquired the ability to analyze any application or scenario on the TRON network. To investigate the basic information of the TRON blockchain, we start by analyzing the fundamental Block information and the basic transaction information within it. As of the writing time of this paper, UTC zone 2024-04-03 22:59:12, the TRON network has reached block number 60,505,000, with a total of 60,505,001 blocks and 8,190,158,591 transactions.

Among the various transaction types, TriggerSmartContract is the most frequent, with a total of 3,437,398,059 transactions, far exceeding other types. This indicates that smart contract calls on the TRON network are very popular among users. Following this are TransferContract and TransferAssetContract, which are transactions for transferring TRX and TRC10 assets. Next are DelegateResourceContract and UndelegateResourceContract, which are transactions for delegating bandwidth and energy resources of accounts. These transactions are evidently sourced from energy leasing services provided by wallets or websites like TokenPocket\footnote{\url{https://help.tokenpocket.pro/en/wallet-faq-en/tron-wallet/energy}} and TRXUS\footnote{\url{https://www.trxus.com}}. Although FreezeBalanceContract and UnfreezeBalanceContract can also provide more transaction energy for accounts, their transaction numbers are significantly lower.

Due to the numerous transaction types in the TRON Protocol, we will specifically explore the ecosystem centered around TRON stablecoins.

\subsection{Decentralization and Development of TRON }

\begin{figure}
\includegraphics[width=0.48\textwidth]{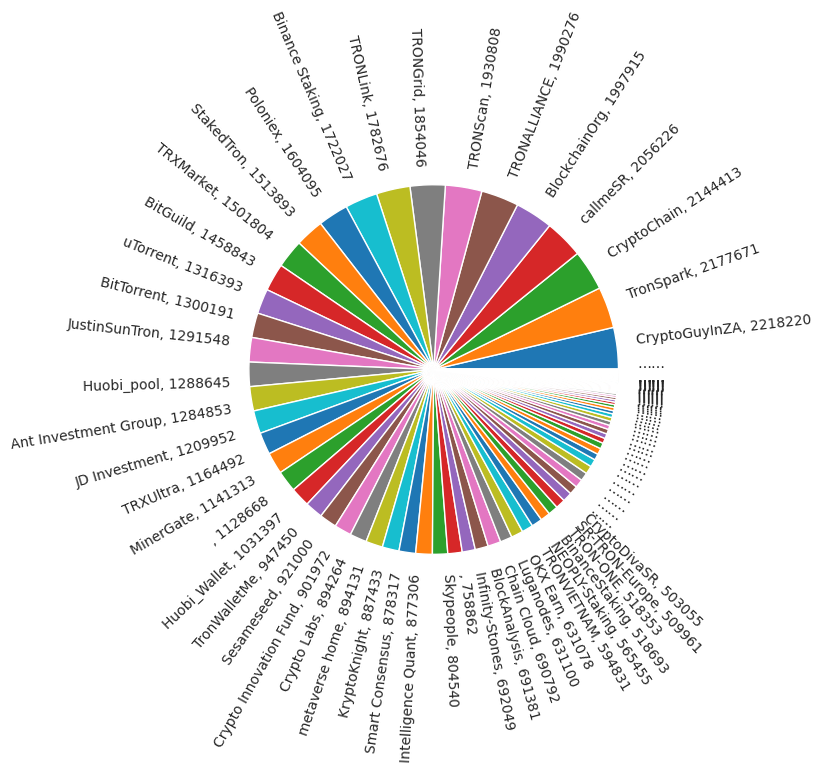}
\includegraphics[width=0.48\textwidth]{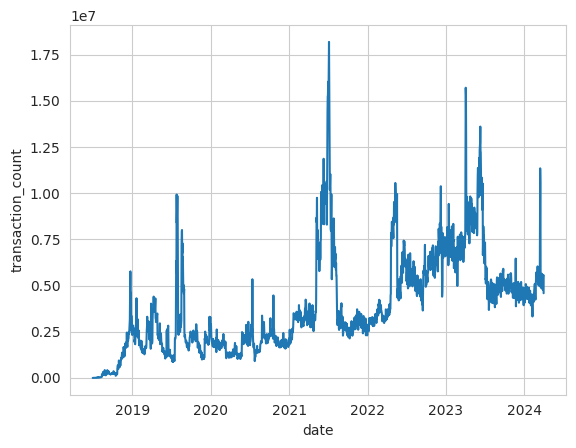}
\caption{The image on the left depicts the distribution of witness addresses across all blocks. The right one illustrates the fluctuation of transaction volume as the block height increases.} \label{fig:blocks_and_txs}
\end{figure}

As mentioned above, TRON utilizes a DPoS consensus mechanism, where witnesses are elected through voting. These witnesses are responsible for generating blocks and maintaining the network and are subject to stringent requirements, including staking a large amount of TRX tokens and continuous community voting supervision. This transparency and accountability of witnesses enable a comprehensive understanding of the network's dynamics, block production efficiency, voting support status, and token flow changes, contributing to the assessment of network security and decentralization. According to \autoref{fig:blocks_and_txs}, despite the relatively small number of witness addresses, the TRON network remains decentralized at the block packaging level, with no single or few dominating witnesses. 

TRON has always touted itself as a high-performance blockchain system with a block time of 3 seconds, significantly faster than Ethereum. However, the actual throughput in real-world scenarios still needs to be analyzed in conjunction with the on-chain block packaging situation. As shown in \autoref{fig:blocks_and_txs}, the daily transaction volume of the TRON network shows an overall upward trend with the increase in block height, but there are fluctuations.

In the early days, TRON's daily transaction volume was relatively low, only a few tens of thousands. As the community ecosystem gradually developed, the number of users and DApps increased, and the transaction volume also gradually grew. By mid-2019, TRON's daily transaction volume had stabilized but had surged to a peak of 1 million at times. Entering 2020, due to the impact of the pandemic on the physical industry, TRON's transaction volume began to grow rapidly, reflecting the network's activity, even reaching a peak of about 18 million in 2021. However, towards the end of 2023, due to an increase in negative news reports about TRON, the transaction volume began to decline sharply, rebounding from the beginning of 2024.

Overall, TRON's transaction volume has gradually increased with the block height, reflecting the continuous development and growth of its ecosystem. Particularly in the past two years, the transaction volume has remained at a high level, indicating that TRON has gained a relatively stable user base and application coverage. However, the fluctuations in transaction volume also suggest that there is still significant room for development in the TRON ecosystem. How to attract more quality projects and funds in the future to ensure ecosystem activity will be a major challenge for TRON. In the long run, the continuous growth of transaction volume is crucial and is a litmus test for TRON's true strength.

\subsection{Chain-predefined Native Services}

\begin{figure}[htbp]
\includegraphics[width=0.48\textwidth]{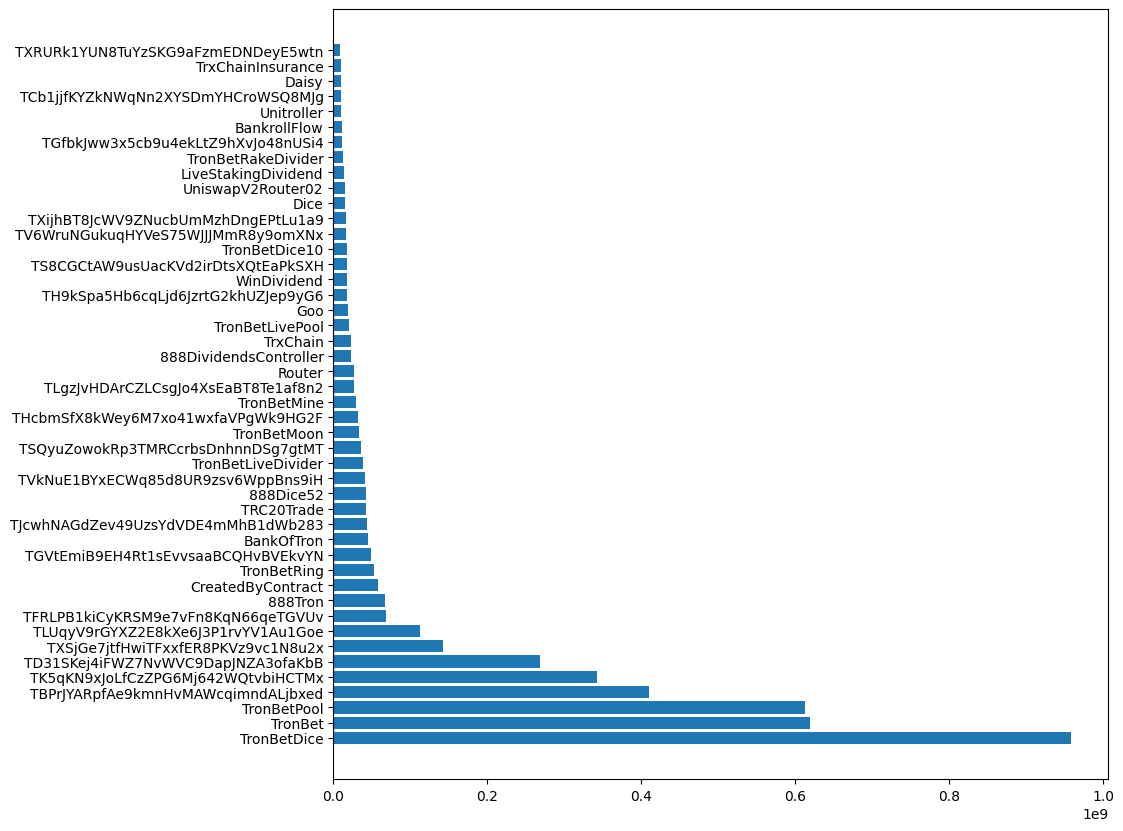}
\includegraphics[width=0.48\textwidth]{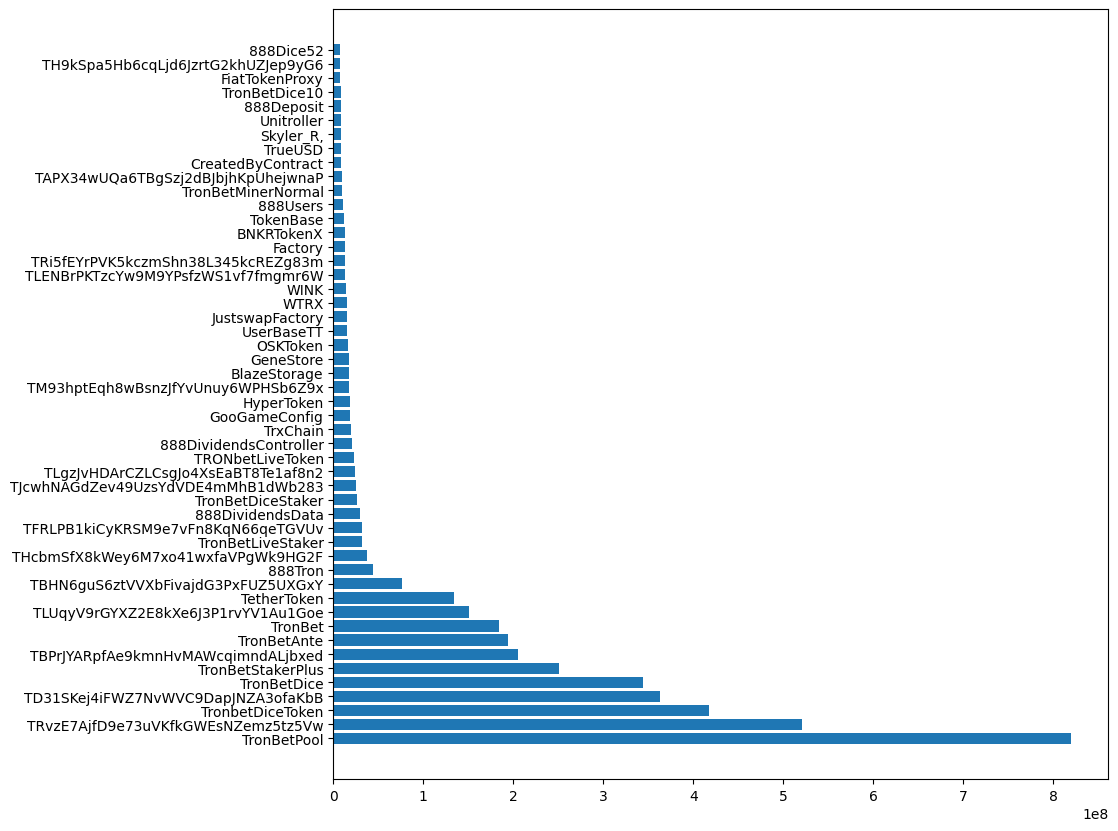}
\caption{The image on the left shows the top 50 sender addresses in terms of the number of internal transactions. The image on the right shows the top 50 recipient addresses in terms of the number of internal transactions.} \label{fig:internals}
\end{figure}

Due to the high degree of flexibility in transactions, TRON natively supports some chain-predefined features like issuing new TRC10 assets and transferring assets.
By analyzing transaction parameters, we can explore these native services on the TRON network.

The most fundamental operation is the TransferContract, which denotes the transfer of TRX tokens. Analyzing the Top 50 list, it is evident that nearly all sender and receiver addresses, regardless of the number of transactions or transaction amounts, belong to centralized exchanges. However, this only represents external transaction information and does not include TRX transfers resulting from contract control. Therefore, further analysis of internal transactions is necessary to explore the actual on-chain scenarios. As shown in \autoref{fig:internals}, the previously mentioned centralized exchange addresses are absent, leaving mostly gambling addresses and addresses of decentralized exchanges like Justswap\footnote{\url{https://justswap.org/}}.

\begin{figure}[htbp]
\includegraphics[width=0.48\textwidth]{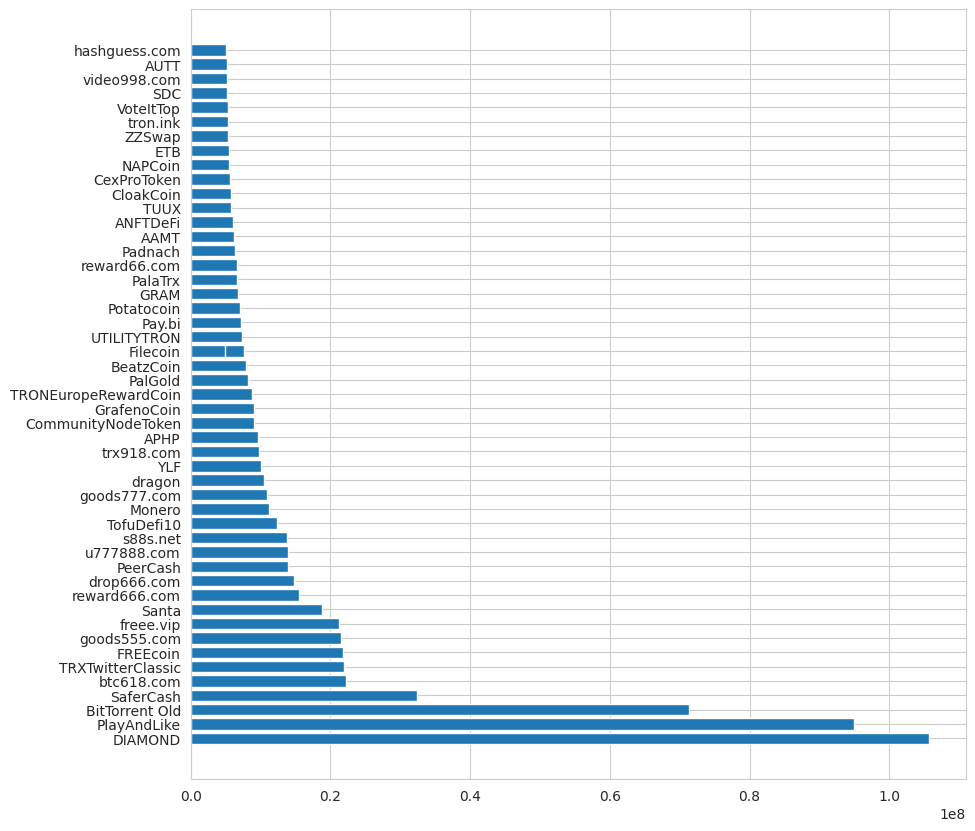}
\includegraphics[width=0.48\textwidth]{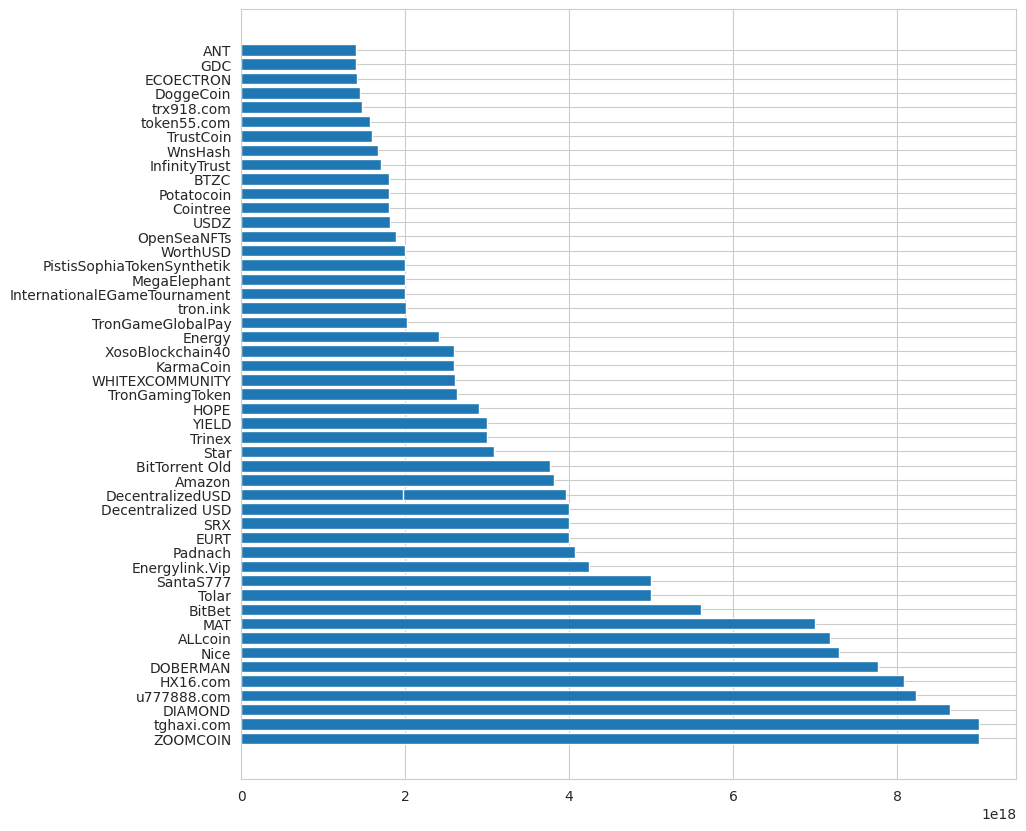}
\caption{The image on the left shows the top 50 TRC10 token names by number of transactions. The right one shows the top 50 TRC10 token names by transaction volume.} \label{fig:trc10}
\end{figure}

The TransferAssetContract represents the transfer of user-created TRC10 tokens. In \autoref{fig:trc10}, we have compiled the number of transactions for various tokens. Interestingly, apart from Justin Sun's Bittorrent token, which is not the most popular, tokens representing transfer services such as Diamond and the blockchain gaming platform token PlayAndLike are more prevalent on the network. Additionally, many token names include gambling website URLs, and further examination of token descriptions revealed a significant amount of promotional information for these gambling sites. However, these gambling sites do not use these tokens as betting currency, and unlike the previously mentioned tokens, these tokens do not have substantial intrinsic value. The high transaction frequency suggests that sending these tokens is used as a promotional strategy.

\begin{figure}[htbp]
\includegraphics[width=0.48\textwidth]{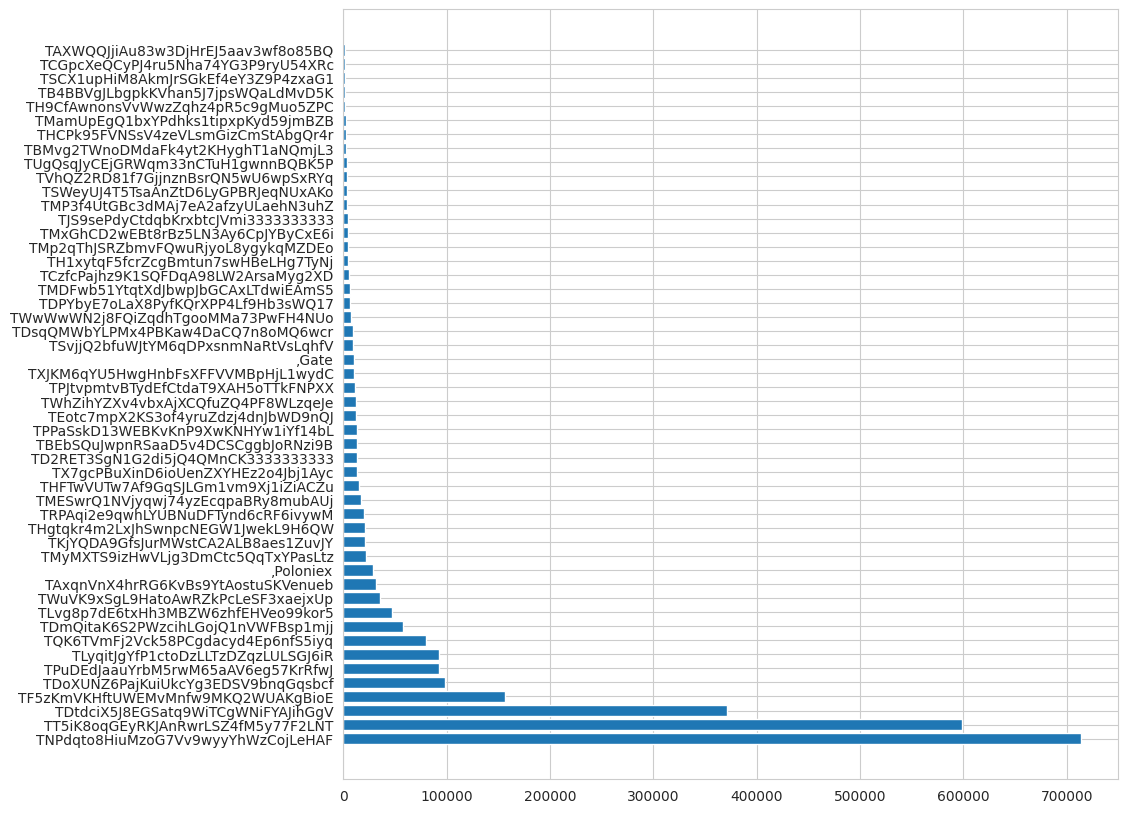}
\includegraphics[width=0.48\textwidth]{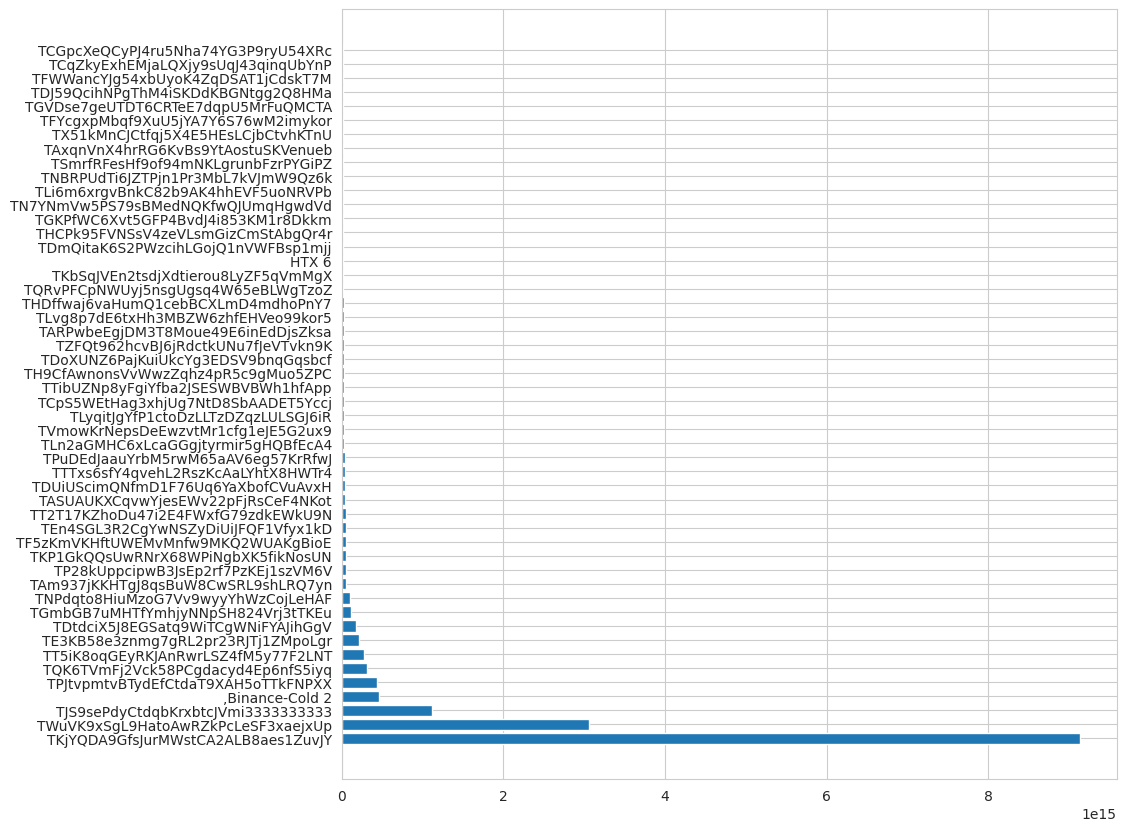}
\caption{The image on the left shows the top 50 addresses in terms of the number of bandwidth delegation occurrences. The right one shows the top 50 addresses in terms of the total bandwidth amount delegated.} \label{fig:delegate_bandwidth}
\end{figure}

\begin{figure}[htbp]
\centering
\includegraphics[width=0.48\textwidth]{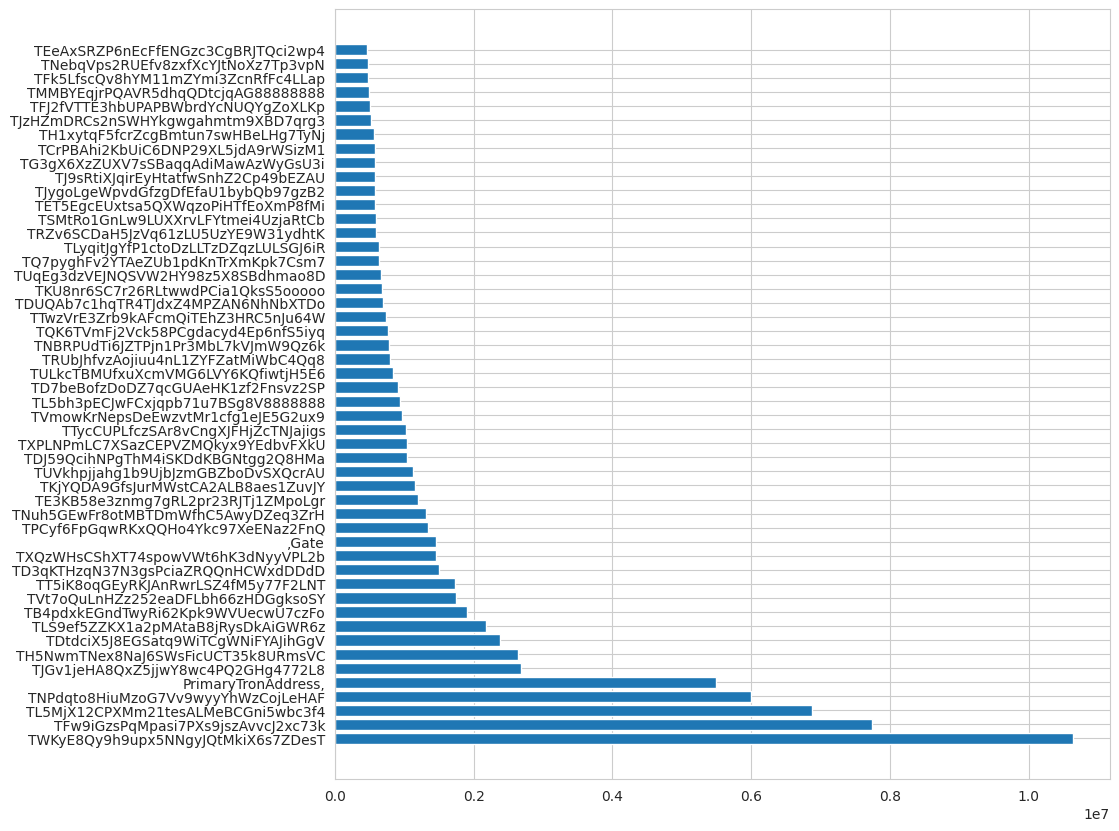}
\includegraphics[width=0.48\textwidth]{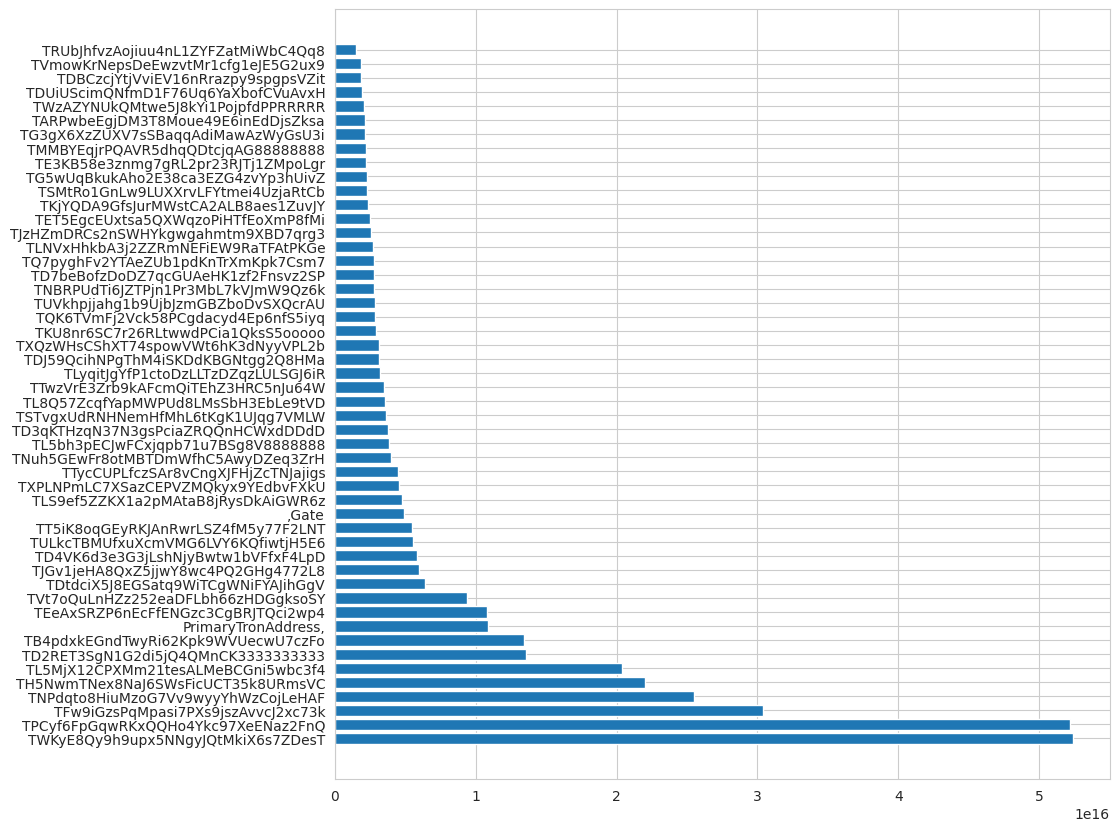}
\caption{The image on the left shows the top 50 addresses in terms of the number of energy delegation occurrences. The right one shows the top 50 addresses in terms of the total energy amount delegated.} \label{fig:delegate_energy}
\end{figure}

Subsequently, we analyzed the Resource Delegate situation. \autoref{fig:delegate_bandwidth} and \autoref{fig:delegate_energy} represent the number and amount of Bandwidth (i.e., assisting others in completing transactions) and Energy (i.e., assisting others in executing smart contracts) delegations, respectively. Although most addresses do not have corresponding labels, it is evident that several service providers specialize in offering resource services to others. Additionally, many centralized exchanges use cold wallets to delegate resources to their hot wallets, thereby reducing on-chain asset transfer costs.

\subsection{Smart Contract and USDT Stablecoin}

\begin{figure}[htbp]
\centering
\includegraphics[width=0.48\textwidth]{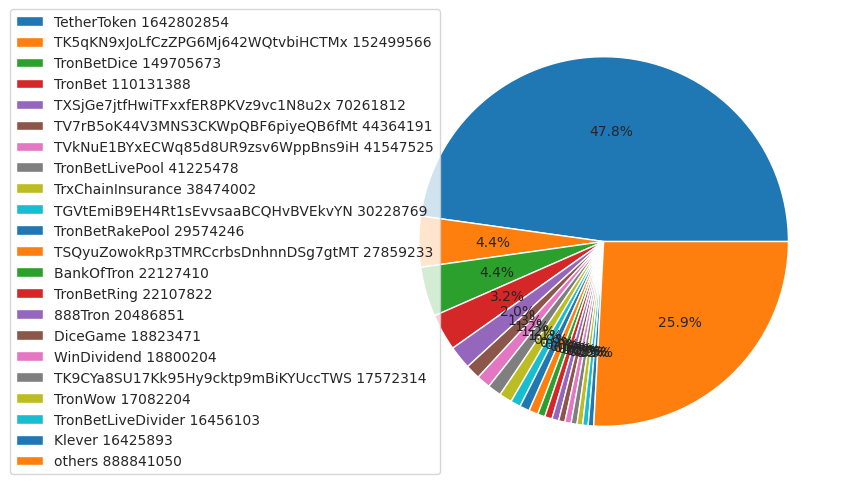}
\includegraphics[width=0.48\textwidth]{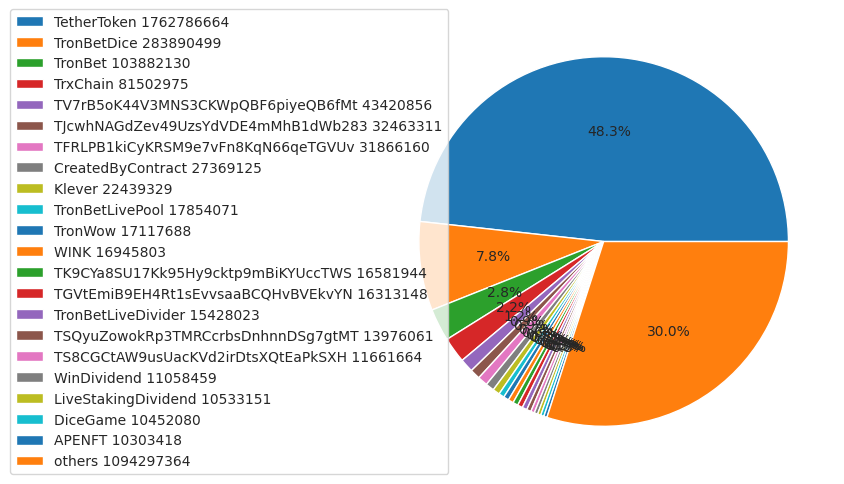}
\caption{The image on the left shows the distribution of addresses triggering smart contract transactions, by the number of occurrences. The image on the right shows the distribution of addresses triggering smart contract events, by the number of occurrences.} \label{fig:smart_contract}
\end{figure}

Unlike the flourishing scene on Ethereum with various tokens and DeFi applications, the statistics of smart contract triggers and event logs on the TRON network are astonishing.

As shown in \autoref{fig:smart_contract}, nearly 50\% of users in TRON who actively trigger smart contract transactions choose to directly operate the TetherToken, i.e., USDT stablecoin. Additionally, from the annotated address types, it can be seen that besides directly operating USDT, users also favor gambling. Smart contracts related to casinos such as TronBetDice, TronBet, 888Tron, DiceGame, and TronWow are quite popular.


We further analyzed the events related to USDT and found that there were 1,750,695,957 Transfer events, 12,089,253 Approval events, 805 AddedBlackList events, 51 RemovedBlackList events, 303 DestroyedBlackFunds events, 257 Issue events for issuing new USDT, and 37 Redeem events. Although we did not delve into these BlackList events this time, we believe our dataset can aid future research on network fund security concerning these BlackLists and related events. We further investigated the more frequent Transfer and Approval events.

From the Transfer events, we established a large list of high-frequency transaction receivers and senders, most of which are cold and hot wallet addresses of centralized exchanges, including Okex, HTX, Binance, Kucoin, MXC, and Bybit. This indicates that centralized exchanges still dominate the use of stablecoins on TRON. Additionally, addresses of well-known high-volume traders like FarFuture also appeared. Similarly, we created a list of the Top 50 USDT holders, which also predominantly consists of centralized exchanges. However, compared to the transaction list, the holder list includes more unlabeled personal or institutional accounts, indicating that many entities trust the security of the TRON blockchain and choose to store large amounts of assets on it.


\subsection{Related Work and Discussion}

Based on our data exploration results, we propose several recommendations for future research on the TRON ecosystem, drawing from existing studies on TRON and other blockchain ecosystems:

\textbf{Market and Audience Analysis of Resource Delegate Services}: While there is technical research on resource management within TRON \cite{li2020resource}, there is a lack of analysis on its real-world application. In contrast, EOS.IO, which also has a complex resource management mechanism, has been the subject of studies analyzing resource usage in real scenarios \cite{zheng2021xblockEOS}. Therefore, we recommend conducting market and audience analyses for the Resource Delegate services in TRON.

\textbf{Real-Time Tracking and Analysis of Hacker Funds on TRON}: Generally, the profits derived from hacker attacks on TRON are frequently channeled through fund splitting or intricate transactions to obscure the paths of illicit money transfers, evading financial regulatory agencies' fund tracking and analysis. While most hacker fund analyses focus on Bitcoin or EVM-based blockchains \cite{goldsmith2020analyzing,tsuchiya2021cryptocurrency}, research on real-time tracking and analysis of hacker funds on TRON is limited. Consequently, this research aims to facilitate the prompt detection of hacker fund transfer addresses, empowering financial regulatory bodies to trace and prevent the circulation of illegal funds. This proactive approach serves to protect investors' funds and uphold the security of the TRON network

\textbf{De-Anonymization of Addresses on TRON}: This research aims to identify and categorize transaction patterns in the TRON ecosystem, with a specific focus on centralized exchanges (CEXs). CEXs act as vital channels for converting cryptocurrencies to fiat currencies and back, playing a key role in enhancing liquidity and accessibility of digital assets \cite{aspris2021decentralized}. Despite the acknowledged importance of CEXs, there is a lack of targeted research on CEXs within the TRON network, with only one existing study focusing on CEXs in the context of Bitcoin \cite{ranshous2017exchange}. While de-anonymization studies have a significant impact on financial regulatory bodies by providing improved tools for regulatory compliance and enhancing their ability to investigate illicit activities on TRON, they also offer valuable insights to investors. By understanding the transaction behaviors of CEXs on TRON, investors can evaluate risk levels associated with different wallet addresses and adjust their investment strategies accordingly.

\textbf{Research on Casino Applications in TRON}: The TRON ecosystem boasts a notable presence of casino applications and associated promotional content
, underscoring the significance of investigating the development and current status of these platforms Existing studies have analyzed online casinos and decentralized casinos on other blockchains, shedding light on the nuances in the game mechanisms, management structures, and player preferences across different platforms \cite{scholten2022behavioural,scholten2020inside,brown2021gambling,meng2020understanding}. Such research holds immense value in shaping the trajectory of gambling entertainment development and facilitating regulatory frameworks within the gambling industry. By delving into the intricacies of casino applications on TRON, researchers can contribute significantly to enhancing transparency, security, and responsible gambling practices within the ecosystem.

\textbf{Study of Stablecoin Illicit Activities on TRON}: The presence of blacklists in on-chain USDT and reports of hacks, money laundering, and terrorist financing on TRON warrant thorough investigation. While there is substantial research on analyzing stablecoin influence on cryptocurrency markets \cite{ante2021influence}, DeFi security \cite{li2022survey}, illicit activities detection  \cite{ibrahim2021illicit,liu2022fa,wu2023know,wu2020phishers} and de-anonymization \cite{huang2022ethereum,wu2021analysis} on major blockchains like Bitcoin and Ethereum, the heterogeneity of data between blockchains means that these detection algorithms may not be directly applicable to TRON. Research aimed at tracing the origin of funds, identifying, and blocking illegal transactions, which is crucial for the compliance of TRON applications and the TRON blockchain itself.

\section{Conclusion}

In summary, this work presents a comprehensive framework for extracting and exploring on-chain data from the TRON blockchain ecosystem. We developed an innovative ETL process tailored specifically to handle TRON's unique data structures and a toolkit for data exploration, we overcame significant technical challenges to acquire large-scale datasets that span all aspects of the TRON blockchain, including blocks, all types of external transaction details, event logs, and internal transactions.

Our in-depth exploration and analysis of the TRON datasets have unveiled novel insights into the blockchain ecosystem. The ecosystem demonstrates reasonable decentralization in block production, indicating a distributed network of validators. However, it is heavily centered around specific types of applications and services. Gambling applications form a significant portion of the ecosystem's activity, highlighting the popularity of blockchain-based gaming and betting platforms. The USDT stablecoin plays a crucial role in facilitating transactions and providing a stable medium of exchange within the network. Additionally, centralized exchanges maintain a strong presence, serving as important gateways between the TRON ecosystem and the broader cryptocurrency market. Our analysis also characterized the resource delegation markets within TRON, shedding light on how computational resources are allocated and traded among network participants. Furthermore, we examined patterns in smart contract usage, providing insights into the types of decentralized applications being built and utilized on the TRON blockchain.

Looking ahead, this foundational dataset establishes a basis for crucial future research into the TRON blockchain. A key area for investigation is the adoption of delegate services, which could reveal important trends in how users interact with and participate in the network's governance. Conducting thorough audits of prominent gambling applications is another priority, as it would enhance our understanding of their operations, fairness, and potential risks. Stablecoin activity on TRON, particularly involving USDT, warrants in-depth examination. This research could uncover patterns of usage and potentially identify any illicit activities associated with stablecoin transactions. In parallel, developing and refining methods to detect and prevent money laundering within the TRON ecosystem is crucial for maintaining the integrity of the network and complying with regulatory standards. The heterogeneous nature of TRON's data opens up exciting possibilities for cross-blockchain research. Exploring transfer learning techniques could allow insights gained from TRON to be applied to other blockchain ecosystems, potentially accelerating research and development across the field. Additionally, developing methods to adapt analysis techniques for use across different blockchain platforms would greatly enhance our ability to conduct comparative studies and identify broader trends in the blockchain space.

By pursuing these research directions, we can deepen the understanding of the TRON ecosystem and contribute valuable insights to the broader field of blockchain analysis. This work has the potential to improve the security, efficiency, and overall functionality of blockchain networks, ultimately driving innovation and adoption in the decentralized technology sector.

\subsection*{Data Availability Statement}

The code used for this research is available in the GitHub repositories mentioned earlier in the footnotes.
The datasets generated and analyzed are too large to be hosted on GitHub. Therefore, they are available at \url{https://web3resear.ch/datasets}.

\bibliography{manual}

\end{document}